\documentclass[12pt,a4paper]{article}
\usepackage[longnamesfirst, authoryear, round]{natbib}

\let\counterwithin\relax
\usepackage[T1]{fontenc}
\usepackage[utf8]{inputenc}
\usepackage[width=1\textwidth, labelfont=bf]{caption}
\usepackage{enumitem}
\usepackage{booktabs}
\usepackage[all]{nowidow} % Unterdrückt Schusterjunge 
\usepackage{amsmath}
\usepackage{amsthm} 
\usepackage{amsfonts}
\usepackage{tikz} % Package to draw tree structures 

\pgfmathdeclarefunction{gauss}{3}{%
  \pgfmathparse{1/(#3*sqrt(2*pi))*exp(-((#1-#2)^2)/(2*#3^2))}%
}

\def\bigsup#1{^{\vbox{\hbox{$\scriptstyle#1$}\nointerlineskip\hbox{}}}}

\usepackage{tikz-qtree} % For tree commands 
\usepackage{pgfplots}
\pgfplotsset{compat = 1.14}
\usepackage{mathtools,amssymb}
\usepackage{xcolor}
\usepackage{cooltooltips}
\usepackage{MnSymbol}
\usepackage{graphicx}
\usepackage{microtype}
\newcommand{\bigCI}{\mathrel{\text{\scalebox{1.07}{$\perp\mkern-10mu\perp$}}}} % Generates the right independence symbol for causal 
\usepackage{subcaption}
\usepackage[flushleft]{threeparttable}

\usepackage{placeins} % Damit die Literatur starr gesetzt ist- Keine Abbildungen können mehr dahin wandern. 
\usepackage{indentfirst} % Macht eine Einrückung nach der Section
\usepackage{lscape}
\usepackage{rotating}
\usepackage[pdfborderstyle={/S/U/W 1},breaklinks,colorlinks]{hyperref}
\AtBeginDocument{%
  \hypersetup{
    citecolor=blue,
    urlcolor =blue,
    menucolor = black,
    linkcolor = black}}
    
\usepackage{tocloft}

\usepackage{chngcntr} %Abbildungs- und Tabellennummer laut Kapiteln
\counterwithin{figure}{section}
\counterwithin{table}{section}

\usepackage{tcolorbox}
\tcbset{colframe=black, colback=white,size=fbox}

\usepackage{algpseudocode}
\usepackage[linesnumbered,ruled]{algorithm2e}

\usepackage[left=2.8cm,right=2.8cm,top=3.5cm,bottom=3.5cm]{geometry}

\usepackage{listings}
\usepackage{xcolor}
\title{Group Average Treatment Effects for Observational Studies\thanks{Financial support from the Deutsche Forschungsgemeinschaft via the IRTG 1792 “High Dimensional Non Stationary Time Series”, Humboldt-Universität zu Berlin, is gratefully acknowledged. All correspondence may be addressed to the author by e-mail at \href{mailto:daniel.jacob@hu-berlin.de}{daniel.jacob@hu-berlin.de}.} \\ }

\date{This Draft: \today}

\author{Daniel Jacob\thanks{ School of Business and Economics, Humboldt-Universität zu Berlin, Spandauer Straße 1, 10178 Berlin, Germany.} 
}

\begin{document}
    \maketitle
\thispagestyle{empty}

\begin{abstract}

%This paper uses survey data from the National Science Foundation (NSF) to investigate heterogeneous effects from treatment. 

The paper proposes an estimator to make inference of heterogeneous treatment effects sorted by impact groups (GATES) for non-randomised experiments. The groups can be understood as a broader aggregation of the conditional average treatment effect (CATE) where the number of groups is set in advance. In economics, this approach is similar to pre-analysis plans. Observational studies are standard in policy evaluation from labour markets, educational surveys and other empirical studies. To control for a potential selection-bias, we implement a doubly-robust estimator in the first stage. We use machine learning methods to learn the conditional mean functions as well as the propensity score. The group average treatment effect is then estimated via a linear projection model. The linear model is easy to interpret, provides p-values and confidence intervals, and limits the danger of finding spurious heterogeneity due to small subgroups in the CATE. To control for confounding in the linear model, we use Neyman-orthogonal moments to partial out the effect that covariates have on both, the treatment assignment and the outcome. The result is a best linear predictor for effect heterogeneity based on impact groups. We find that our proposed method has lower absolute errors as well as smaller bias than the benchmark doubly-robust estimator. 
We further introduce a bagging type averaging for the CATE function for each observation to avoid biases through sample splitting. The advantage of the proposed method is a robust linear estimation of heterogeneous group treatment effects in observational studies. 	

%Since flexibility in terms of the model choice, as well as interpretability of the results, is of main interest
%These findings show high potential for policy decision making among faculties. 

%The estimation of a causal parameter in a high-dimensional setting where the functions are potentially complex is a challenging task. Parametric and linear modelling is often not sufficient to generate unbiased and consistent estimators. Modern approaches, therefore, use machine learning algorithms to learn these nuisance functions. However, this leads to new problems like the regularization bias or overfitting that are common when using such models.


\vspace{1.0cm}

\end{abstract}

\newpage
\setcounter{page}{1}

\section{Introduction}

%To investigate this causal effect we use survey data from the National Science Foundation (NSF). The current dataset is from 2017 and consists of around 85,000 observations and more than 200 covariates. 
%Different from these papers we do not focus on a descriptive analysis rather than find the best function for heterogeneous treatment effects and then look at which variables are drivers for such differences. 

%To the best of our knowledge, this is the first time that this question is investigated. 

%we are not aware of any other paper that tries to answer this question.

When evaluating the causal effect of some policy, marketing action or other treatment indicator, it might not be sufficient to report the average treatment effect (ATE). The estimation of heterogeneous effects, e.g. the conditional (on covariates) average treatment effect (CATE), provides further insight into causal mechanisms and helps researchers and practitioners to actively adjust the treatment assignment towards an efficient allocation. The more information in terms of characteristics is provided, the better can heterogeneity be observed. If we have little deterministic information, it might be that heterogeneity effects are overlooked. The trade-off here is that the more covariates datasets have, the more complex they get. We might not know the structural form of our functions of interest or we simply have more covariates than observations. Even if the number of raw covariates is smaller than the number of observations, including quadratic or even higher order interaction increases the number of covariates and can easily exceed the sample size. This is why parametric models are often insufficient when applied on high-dimensional, non-linear datasets \citet*{chernozhukov2018double}.

Under the assumption of approximate sparsity, meaning that out of our rich covariate space only a few are dependent on our variables of interest (the outcome and the treatment assignment), recent methods for treatment effect estimation use machine learning models that have shown to be superior in high-dimensional prediction problems \citet*{hastie2009elements}. The idea is to learn nuisance functions and regularize the parameter space while making as little assumptions as possible. This is especially helpful when the data does not come from randomised experiments where treatment is randomly assigned to the individuals. In observational studies, self-selection into treatment can arise and introduce a bias that has to be corrected for (i.e. self-selection bias) \citep{heckman1998characterizing}. For the ATE one would use the nuisance parameter to orthogonalize the effect that covariates have on both, the treatment assignment and the outcome variable. See \citet{chernozhukov2018double} for a recent approach, which they call double machine learning.  

Methods to estimate the CATE are, among others, the generalized random forest, which builds on the idea of controlling for observed confounders through a tree structure and then estimates the CATE within each final leaf \citep{athey2019generalized}. Another approach is causal boosting, which uses the idea of causal trees as a weak learner \citep{powers2018methods}. What the aforementioned methods lack, however, is that they are built on tree algorithms and therefore do not allow a flexible estimation of heterogeneous treatment effects in terms of the underlying algorithm (e.g. LASSO or Neural Networks). A recent method called R-learner does provide such flexibility and shows competitive performance in the estimation of the CATE \citep{nie2017quasi}. Other models, known as meta-learners, decompose the modelling procedure into sub-regression functions, which can be solved using any supervised learning method. This can, for example, be done by a two-model approach (TMA) where a response function (conditional mean) on the treated and another one on the non-treated observations is estimated. The difference between the two functions can thus be interpreted as the CATE \citep{kunzel2019metalearners}. This approach falls into the category of indirect estimation since its goal is to minimize the squared error loss based on the outcome variable.  A more efficient approach is to directly estimate the treatment effect and regularize based on the treatment effect itself rather than the observed outcome. This leads to a reduced variance of the prediction \citep{hitsch2018heterogeneous}. In order to do this, we first have to estimate a proxy function of the treatment effect. The literature refers to this approach as the treatment effect projection \citep{hitsch2018heterogeneous} or the modified outcome method \citep{knaus2018machine}. The idea is to use inverse probability weighting or the doubly-robust estimator as proposed by \citep{robins1995semiparametric}. Using the estimates from the two-model approach in combination with inverse probability weighting (IPW) decreases the variance of the estimator and controls for observed confounding (see e.g. \cite{lunceford2004stratification}). Additional orthogonalization using the two conditional mean functions produced by the TMA further decreases the bias of the parameter of interest \citep{lee2017doubly}. The doubly-robust estimator can be used in high-dimensional settings to estimate a reduced dimensional conditional average treatment effect function. Functional limit theory can be derived for the case where the nuisance functions are trained via machine learning methods, which are then applied on the doubly-robust estimator \citep{fan2019estimation,zimmert2019group}. Recent papers study and evaluate different models that are designed for the estimation of heterogeneous treatment effects (see e.g. \cite{knaus2020double, knaus2018machine, kunzel2019metalearners, powers2018methods}. 

The difficulty, however, is that machine learning methods are often a black box that is not easy to interpret. This fact hinders the information on drivers for effect heterogeneity. In this paper, we, therefore, build on the ideas of \citet{chernozhukov2018generic}, who concentrate on estimating group average treatment effects (GATE) in randomised experiments. The groups are built on the distribution from the CATE (e.g. quantiles to get five groups). A semi-parametric model is then used to identify the best linear predictor for the group treatment effect, providing standard errors and confidence intervals. The heterogeneity between the groups can further be interpreted through covariates, which shed some light on the question of what characteristics determine the differences between groups. In this paper, we extend the approach to estimating the GATE parameter towards the use in observational studies and also towards the possibility to estimate a best linear CATE based on group heterogeneity. In a paper that is very close to ours the authors use a linear model with the aforementioned orthogonalization step to estimate group treatment effects based on observed covariates \citep{park2019groupwise}. They define the groups by conditioning on observed covariates, for example, for a binary variable they estimate the average treatment effect for the two levels of the variable.  

The advantage of the method proposed here is a robust estimation of group heterogeneous treatment effects that is comparable with other models thus keeping its flexibility in the choice of machine learning methods and at the same time its ability to interpret the results. The latter is especially useful in all areas of empirical economics like policy or labour markets interventions. It also has the advantage to control for potential self-selection bias. The idea of going beyond the average, but not as deep as to estimate conditional average treatment effects for many covariates, is first considered in \citet{chernozhukov2018sorted}. They provide standard errors and confidence bands for the estimated sorted group effects and related classification analysis and provide confidence sets for the most and least affected groups. While they only use parametric estimators, a non-parametric attempt to estimate group average treatment effects and also provide insights from the heterogeneity in terms of observed covariates is proposed by \citet{fan2019estimation} and \cite{zimmert2019group}. They use a two-step estimator of which the second step consists of a kernel estimator. 

Our contribution is to use the doubly-robust estimator and thus keep the flexibility in using any kind of machine learning method to learn the nuisance parameter in the first step. In a second step, we use Neyman-orthogonal scores to cancel out the effect that covariates can have on both, the outcome and the treatment selection. Using the residuals, we set up a linear model to learn the group average treatment effects.  When grouping the heterogeneous treatment effect, we find that we can get more precise estimates in terms of mean absolute error (MAE) and Bias with our proposed method. Especially the linearization increases the accuracy of our estimates. We also show that the CATE function, which we estimate via a doubly-robust estimator, should be weighted over multiple iterations. This type of bagging further decreases the MAE of the CATE function. We split our paper in three main parts. First, we state the methodology for randomized experiments and second, the extensions for observational studies. Third, we employ a extensive simulation study that include selection bias, high-dimensionality and non-linearity in the data generating process.

\section{Generic Machine Learning for Group ATE}

\subsection{Potential Outcome Assumptions}

Throughout this paper, we make use of the potential outcome theorem \citep{rosenbaum1983central} and state four necessary assumptions.
The first assumption is the ignorability of treatment, conditional on observed covariates ($X$), from the two potential outcomes. It is also known as unconfoundedness or simply conditional independence: 

\begin{align}
\left(Y_{i}^{1}, Y_{i}^{0}\right) \bigCI D_{i}|X_{i}. \label{PotOut}
\end{align}
With $Y^1$ denoting the potential outcome under treatment and $Y^0$ if not being treated. $D$ is the treatment assignment variable. 

The second assumption, the Stable Unit Treatment Value Assumption (SUTVA), guarantees that the potential outcome of an individual is unaffected by changes in the treatment assignment of others. This assumption might be violated if individuals can interact with each other (peer and social effects).

The third assumption, called overlap, guarantees that for all $x \in supp(X)$,  the probability of being in the treatment group (i.e. the propensity score, $e_{0}(X)$), is bounded away from 0 and 1:

\begin{align}
\quad 0 < \operatorname{P}(D=1|X=x) < 1. \nonumber \\
e_{0}(X) = \operatorname{P}(D=1|X=x). 
\end{align}

We control for the common support by estimating the propensity score and balance the treatment and control group based on the distribution. We hence exclude all observations that have a propensity score lower $0.02$ or higher than $0.95$.
The fundamental problem of causal inference is that we only observe one of the two potential outcomes at the same time. The counterfactual for a nontreated (treated) person, namely, what would have happened if this person were (not) treated, is always missing. We can represent this statement through a switching regression where the observed outcome ($Y_i$) depends on the two potential outcomes and the treatment assignment: 

\begin{align}
    Y_i = Y_i^0 + D(Y_i^1-Y_i^0).
\end{align}

We further assume that, for the estimation of standard errors, the following moments exist: $\operatorname{E}\left[|Y^j|^q\right] < \infty$ for $q \geq 4$ and $j =0,1$.

\subsection{Randomized Control Trial} \label{Sec:RCT}

To provide valid estimation and inference for a causal interpretation of parameters, \citet{chernozhukov2018generic} focus on features of the CATE. One of the main features is the \textbf{Sorted Group Average Treatment Effect (GATES)}. The idea is to find groups of observations depending on the estimated treatment effect heterogeneity.  Their proposed method relies on a two-model approach in the first step. Here, two response functions are trained separately for the treated and non-treated observations. This approach will be biased if the data sample is from an observational study. In randomized control trials, the difference between the two functions provides an estimate of the treatment effect for every observation. We refer to $S(X)$ as our proxy-predictor: 

\begin{align}
\tau(X) &= \operatorname{E}[Y|D=1,X] - \operatorname{E}[Y|D=0,X],  \label{tau_expectation} \\  
    {S}(X) &= {g}_{1}\left(X\right)-{g}_{0}\left(X\right). \nonumber
\end{align}

\begin{figure}[ht]
\begin{center}
\begin{tikzpicture}
\begin{axis}[
  no markers, 
  domain=0:6, 
  samples=100,
  ymin=0,
  axis lines*=left, 
  xlabel=$\tilde{S}(X)$,
  every axis y label/.style={at=(current axis.above origin),anchor=south},
  every axis x label/.style={at=(current axis.right of origin),anchor=west},
  height=5cm, 
  width=12cm,
  xtick=\empty, 
  ytick=\empty,
  enlargelimits=false, 
  clip=false, 
  axis on top,
  grid = major,
  hide y axis
  ]

 \addplot [very thick,cyan!50!black] {gauss(x, 3, 1)};

\pgfmathsetmacro\valueA{gauss(2,3,1)}
\pgfmathsetmacro\valueB{gauss(2.8,3,1)}
\draw [gray] (axis cs:2,0) -- (axis cs:2,\valueA)
    (axis cs:4,0) -- (axis cs:4,\valueA);
\draw [gray] (axis cs:2.8,0) -- (axis cs:2.8,\valueB)
    (axis cs:3.2,0) -- (axis cs:3.2,\valueB);

\node[below] at (axis cs:1.3, 0)  {$\gamma_1$}; 
\node[below] at (axis cs:2.4, 0)  {$\gamma_2$}; 
\node[below] at (axis cs:3, 0)  {$\gamma_3$};
\node[below] at (axis cs:3.6, 0)  {$\gamma_4$}; 
\node[below] at (axis cs:4.7, 0)  {$\gamma_5$};
%\node[below] at (axis cs:3.32,0.47) {$\bar{S}(X)$};
\end{axis}

\end{tikzpicture}
\end{center}
\caption{Illustration of a treatment effect function and how we define the GATE.}
\label{density_example}
\end{figure}
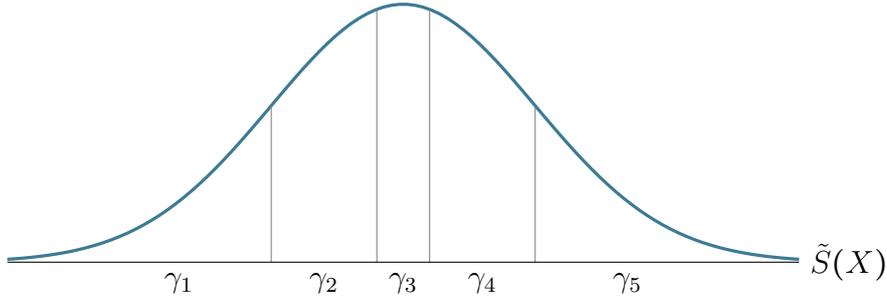

Here $ {g}_{D}\left(X\right)= \operatorname{E}(Y|D,X)$ is the regression model of the outcome variable on $X$ separately for $D \in \{0, 1\}$. The two functions can be estimated with a broad range of supervised machine learning methods. The target parameters are 

\begin{align}
\operatorname{E}[\tau(X)|G_{k}] && G_{k}:= \{S(X) \in I_k\}, \quad k = 1,...,K,
\end{align}

where $G$ is an indicator of a group membership with $I_k = [\ell_{ k-1},\ell_k)$ and $\ell_k$ is the $k/K$-quantile of $\{{S}_i\}_{i \in M}$. Subscript $M$ denotes that these are all out-of-sample predictions. We will for readability not always refer to the sets but always make use of sample splitting when using machine learning methods. 
The groups are ex-post defined by the predicted score function in the first stage. Figure \ref{density_example} shows an illustration of how the GATE parameter are defined.

 If the treatment effect for the groups are consistent, it asymptotically holds that

\begin{align}
\operatorname{E}[\tau(X)|G_{1}] \leqslant \operatorname{E}[\tau(X)|G_{2}] \leqslant ...\leqslant \operatorname{E}[\tau(X)|G_{k}],
\end{align}

which is the monotonicity restriction. Furthermore, it can be tested whether there is a homogeneous effect if $\operatorname{E}[\tau(X)|G_{k}]$ would be equal for all $k$ groups. 
The weighted linear projection equation to recover the GATES parameter is: 

\begin{align}
YH = {\beta}^{\top} A_{1}H + \sum_{k=1}^{K} {\gamma_k} \cdot \mathbf{I}({S}(X) \in I_k) + \nu, \label{equ:GATES}
\end{align}

with
%$\operatorname{E}[w(X) \nu W]=0$,
 $A_{1} = (1,B(X))$ and 
 %$W=(X_{1}',W_{2}')$ with $W_{2}=[\{(D-p(X))1(G_{k})\}_{k=1}^{K}]'$. \\
$B(X) = \operatorname{E}[Y|D=0,X]$ being the baseline function without treatment. 
See pseudo-code of Algorithm \ref{pseudo:1}, which describes the implementation of this method. The weights $H$ represent the Horvitz-Thompson transformation \citep{horvitz1952generalization}:

\begin{align}
H = H(D,X) = \frac{D-{e}(X)}{{e}(X)(1-{e}(X))}. 
\end{align}

This estimator, which is applied to account for different proportions of observations within strata in a target population, is equivalent to the simple inverse probability weighting estimator. These estimators, however, might exhibit a high variance if the identification (the precision) of the propensity scores is lacking \citep{lunceford2004stratification}.  

%The parameter $\gamma$ is the sorted group average treatment effect for each group $1...K$. 

%where $X_{1} = [1,B(X)]$ and $W=(X_{1}',W_{2}')$ with $W_{2}=(\{(D-p(X))1(G_{k})\}_{k=1}^{K})'$.
%$B(X) = E[Y|D=0,X]$ is the baseline function without treatment and the projection $\tau(X) = E[Y|D=1,X] - E[Y|D=0,X]$ is the treatment effect \citep{chernozhukov2018generic}. 
%The parameter $\gamma$ is the sorted group average treatment effect for each group $1...K$. 
The main identification result is that the projection coefficients $\gamma_k$ can be represented in the following way: 

\begin{align}
\gamma = (\gamma)_{k=1}^{K} = (\operatorname{E}[\tau(X) | G_{k}])_{k=1}^{K} \label{equ:gamma}.
\end{align}

\vspace{5mm}

\begin{algorithm}[H] \label{pseudo:1}
    %\SetKwInOut{Input}{Input}
%\small
    
    \For{b=1 to B}{
    \textbf{Split} Data in $k=2$ samples: $I^{a}$ and $M$ with $I^{a} \cupdot M$ \\
    \textbf{Train} $Y_{i}^0 =   g_{0}(X_{i},D=0) + U_{0i}$, with $i \in I^{a}$ \\
     \textbf{Train} $Y_{i}^1 =   g_{1}(X_{i},D=1) + U_{1i}$, with $i \in I^{a}$ \\
      \hskip 1.0cm \textbf{Predict} $\hat{Y}_{i}^{0} = \hat{g_{0}}(X_{i})$, with $i \in M$ \\
      \hskip 1.0cm  \textbf{Predict} $\hat{Y}_{i}^{1} = \hat{g_{1}}(X_{i})$, with $i \in M$ \\ 
        \hskip 2.0cm \textbf{Calculate} $S_{b}(X|i)$ = $\hat{Y}_{i}^{1}-\hat{Y}_{i}^{0}$ \\
      
      \textbf{Train} $D_{i} = e_{0}(X_{i}) + V_i $, with $i \in I^a$ \\
        \hskip 1.0cm \textbf{Predict} $\hat{D_{i}} = \hat{e}(X_{i})$, with $i \in M$ \\
        \hskip 2.0cm \textbf{Calculate} $\hat{V}_{i} = D_{i} - \hat{e}(X_i)$, with $i \in M$ \\
      \textbf{Estimate} GATES parameters ($\gamma$) with weighted OLS using $M$ (see equation \ref{equ:GATES}) \\ 
      
        %\hskip 2.0cm \textbf{Store} $S_{b}(Z|i)$ for b = 1 to B

      }
      \textbf{Average} $\gamma$ over $B$ iterations: $\Tilde{\gamma} = median\{\gamma\}$\\
    \caption{GATES}
\end{algorithm}
\noindent
{\footnotesize The algorithm is based on \cite{chernozhukov2018generic}.}

\vskip 0.5cm

There are two potential sources of uncertainty. One is estimation uncertainty regarding our parameter of interest, keeping sample splitting fixed. The second source is exactly due to the sample splitting. 
To account for this, the p-values, as well as the confidence intervals, need to be adjusted. \citep{chernozhukov2018generic} show, that sample-splitting-adjusted p-values can have the following form.

\begin{align}
\mathbb{P}(p_A \leq \alpha/2 | \text{Data}) \geq 1/2,
\end{align}

with  $p_A$ being the realized p-value given the auxiliary sample and $\alpha$ is the significance level. Given that we use medians to average the parameters $\gamma_k$ over $B$ bootstrap repetitions it holds that for at least 50\% of the random data splits out of $B$, the p-value is at largest $\alpha/2$. Small values provide evidence that the group parameter is different from zero. 
\subsection{Extensions for Observational Studies}

To use the best linear predictor for group heterogeneity in observational studies, we change and extend the TMA in the first step and the linear model in the second step. First, we replace the two-model approach by a doubly-robust estimator. This means we not only weight by the inverse of the propensity score but also orthogonalize the outcome variable by subtracting the conditional mean. We also make use of sample splitting as a form of cross-fitting. The auxiliary sample is applied to estimate the score function via the doubly-robust estimator and the main sample to predict the final score function, which is used in the linear step. In this way, we limit the danger of overfitting. 
%Second, we use medians of the score-functions over multiple iterations to account for potential bias due to sample splitting. 
The resulting function is a more robust version of the CATE for each individual as well as for the GATE function. The two steps are described in more detail in the following.

The function in equation \ref{equ:dr_model} is calculated using the training data (the $I^a$ sample). In a second step, a new supervised model is trained on the transformed outcome using  $I^a$  while predictions are made on the test set $M$ to get an unbiased estimate (see equation \ref{equ:dr_model}. Algorithm \ref{pseudo:2} describes this process.

\begin{align}
\hat{S}_{i} &= \hat{g}_{1}\left(X_{i} \right)-\hat{g}_{0}\left(X_{i} \right)+\frac{D_{i}\left(Y_{i}-\hat{g}_{1}\left(X_{i} \right)\right)}{\hat{e}\left(X_{i}\right)}-\frac{\left(1-D_{i}\right)\left(Y_{i}-\hat{g}_{0}\left(X_{i} \right)\right)}{\left(1-\hat{e}\left(X_{i}\right)\right)} \label{equ:dr_calculation} \\
\hat{S}_{i} &= l(X_i) + \omega \label{equ:dr_model}
\end{align}

 In equation \ref{equ:dr_calculation},  $\hat{g}_{1}\left(X_{i} \right)-\hat{g}_{0}\left(X_{i} \right)$ is equivalent to the score-function from the two-model approach. Simulation evidence from \citet{knaus2018machine, powers2018methods} suggests that estimators based on $\hat{S}_{i} $ might be more stable because of the doubly-robust property and that the performance is competitive for the estimation of heterogeneous treatment effects in observational studies. The doubly-robust property states that the estimator is consistent and unbiased if only one of the models, the regression or the propensity score, is correctly specified \citep{robins1994estimation, robins1995semiparametric}. \citet{lunceford2004stratification, williamson2014variance, belloni2014inference} study the theoretical properties of the doubly-robust estimator and highlight implications for practice. One of the findings is that the variance can be decreased when using the doubly-robust estimator instead of a simple inverse probability estimator \citep{lunceford2004stratification}. \citet{vira2017} show that equation \ref{equ:dr_calculation} is conditionally locally robust to the estimation error of the nuisance parameter.
 
Next we state some asymptotic results to recover the CATE. From equation \ref{tau_expectation} it follows that 
\begin{align}
\tau(X) &= \operatorname{E}\left\{\operatorname{E}[Y|D=1,X] - \operatorname{E}[Y|D=0,X] | X = x_i \right\}  
\end{align}
 
Let  $\eta(X) := ({e}(X), {g}_{1}\left(X_{i} \right),{g}_{0}\left(X_{i} \right))$ be the true high dimensional nuisance parameters. Following \citet{fan2019estimation} we can define 

\begin{align}
\psi(D,Y,X,\eta(X)) = &{g}_{1}\left(X_{i} \right)- {g}_{0}\left(X_{i} \right) \nonumber \\ 
&+\frac{D_{i}\left(Y_{i}-{g}_{1}\left(X_{i} \right)\right)}{ {e_{0}}\left(X_{i}\right)}-\frac{\left(1-D_{i}\right)\left(Y_{i}- {g}_{0}\left(X_{i} \right)\right)}{\left(1- {e_{0}}\left(X_{i}\right)\right)}
\end{align}.

\textbf{Theorem 2.1}  \\
(i) under Assumptions 1,2,3,4 

\begin{align*}
\operatorname{E}&\left[{g}_{1}\left(X_{i} \right)+\frac{D_{i}\left(Y_{i}-{g}_{1}\left(X_{i} \right)\right)}{ {e_{0}}\left(X_{i}\right)} | X = x_i \right] = \operatorname{E}\left[Y^1 | X = x_i \right] , \\
\operatorname{E}&\left[{g}_{0}\left(X_{i} \right)+\frac{(1-D_{i})\left(Y_{i}-{g}_{0}\left(X_{i} \right)\right)}{ 1-{e_{0}}\left(X_{i}\right)} | X = x_i \right] = \operatorname{E}\left[Y^0 | X = x_i \right] 
\end{align*} \\
\noindent
(ii)$\operatorname{E}\left[\psi(D,Y,X,\eta(X)) - \tau(X) | X = x_i \right] = 0$ given (i). \\

This moment condition satisfies the Neyman-orthogonality condition. Neyman-orthogonality is a key component in ensuring that the CATE estimators are robust to the regularization bias inherent for the nuisance functions, which are learned via machine learning models. \\

Next, we set up a linear model for the estimation of the low-dimensional parameters of interest $\gamma_k$. 
Suppose that the data generating process follows a partial linear model that has the following form:

\begin{align}
Y = \tau(X)&D + \mu_{0}(X) + U,  &&E[U | X,D] = 0, \\
&D = e_{0}(X) + V,  &&E[V | X] = 0,  \\
\operatorname{E}[\tau(X)] &= \frac{1}{K}\sum_{k=1}^{K}\gamma_{k}(X).
\end{align}
We define $\gamma$ as before as $\gamma = (\gamma)_{k=1}^{K} = (\operatorname{E}[\tau(X) | G_{k}])_{k=1}^{K}$.
The outcome variable $Y$ depends not only on the treatment effect parameter but also on observed covariates through the function $\mu_{0}(X)$. The second equation displays the setting in observational studies, namely that the treatment assignment also depends on observed covariates through the function $e_{0}(X)$. In randomized control trials, it is sufficient to directly learn the regression function  $\hat{\tau}(X)D$. We do not need to include the function $\mu_{0}(X)$ since in RCT the distribution of the covariates are assumed to be the same for both, the control and the treatment group. The linear model in section \ref{Sec:RCT} does exactly this but also takes into account that the treatment assignment might be different for strata in the covariate space.

In observational settings, confounding through covariates leads to a bias that we have to account for. We wish to partial out the effect from $X$ on $D$ as well as the effect from $X$ on $Y$. Following \cite{chernozhukov2018double}, we again make use of Neyman-orthogonal moments, which leads to the orthogonalized regressors $\hat{U} = Y-\hat{\mu}(X)$  and  $\hat{V} = D-\hat{e}(X)$. The terms $\hat{U}$ and $\hat{V}$ are the residuals that we use in the linear projection function in Equation \ref{equ:GATES_dr}. The propensity-score estimates ($\hat{e}(X)$) can be derived by using the main sample on the already estimated propensity-score function, which is used in the doubly-robust step. The function $ \hat{\mu}(X)$ is estimated using any machine learning model on the auxiliary sample. For the estimation of the average treatment effect $\hat{\tau}$, the residualized regression function has the following form: 
 
 \begin{align}
\hat{\tau} = \left(\frac{1}{N}\sum\limits_{i \in M}\hat{V}_{i}\hat{V}_{i}\right)^{-1} \frac{1}{N}\sum\limits_{i \in M} \hat{V}_{i}(Y_{i}-\hat{\mu}(X_{i})).
\end{align}
 
In our case, we are not specifically interested in the average treatment effect but the average effect given the quantiles of the CATE function. This leads to the linear projection equation to estimate the group average treatment effect using the main sample of observations:

\begin{align}
(Y-\mu_{0}(X))=  \sum_{k=1}^{K}\gamma_{k} \cdot (D-e_{0}(X)) \cdot \mathbf{I}({S}(X) \in I_k) + \nu. \label{equ:GATES_dr}
\end{align}

Let $\tilde{S}(X) = \hat{l}(X)$ and rewrite the empirical analog for the $k$-th specific group as: 

\begin{align}
\hat{\gamma}_k = \left(\frac{1}{N}\sum_{i \in M}\hat{V}_{i} \hat{V}_i \cdot  \mathbf{I}(\tilde{S}(X_i) \in I_k)\right)^{-1} 
\frac{1}{N} \sum_{i \in M}\hat{V}_i\left(Y_{i}-\hat{\mu}(X_i)\right) \cdot  \mathbf{I}(\tilde{S}(X_i) \in I_k).
\end{align}

Through orthogonalization, we can overcome the regularization bias, present when naively employing ML models to estimate the parameter $\gamma$ in this setting. Note that we can use different ML algorithms for each function. A trade-off to regularization bias is overfitting, which can be taken care of through sample splitting. However, through sample splitting, we can only use $N-n$ observations, given that $n$ observations are in the training data. To account for the uncertainty through sample splitting, we run multiple repetitions via bootstrapping and split the sample randomly in each bootstrap. We average the GATES parameter by taking the median over $B$ repetitions.

Naturally, we can do the same sort of bagging for the score-function. Given that we split our sample into two equal parts (one for training and one for estimation), we get estimates of the CATE for half of the observations in each iteration. If we repeat this procedure often enough, we get estimates for every observation and decrease the uncertainty bias due to sample splitting. We then take medians for each observation to get the final CATE estimate ($\bar{S}_i(X)$): 

\begin{align}
\bar{S}_i(X) = median\{\hat{\hat{S}}_b(X)\}.
\end{align}

In our simulation study we show exactly how bagging for CATE can influence the results.

Algorithm \ref{pseudo:2} shows the steps to identify the group treatment effect and the steps to average the CATE function over all repetitions.

%Algorithm \ref{pseudo:2} shows the pseudo-code which includes the two extensions. 
\vspace{5mm}

\begin{algorithm}[H] \label{pseudo:2}
    %\SetKwInOut{Input}{Input}
%\small
    
    \For{b=1 to B}{
    \textbf{Split} Data in $k=2$ samples: $I^{a}$ and $M$ with $I^{a} \cupdot M$ \\
    \textbf{Train} $Y_{i}^0 =   g_{0}(X_{i},D=0) + U_{0i}$, with $i \in I^{a}$ \\
    \textbf{Train} $Y_{i}^1 =   g_{1}(X_{i},D=1) + U_{1i}$, with $i \in I^{a}$ \\
    \textbf{Train} $D_{i} = e_{0}(X_{i}) + V_i $, with $i \in I^a$ \\
    \textbf{Train} $Y_i = \mu(X_i) + U_i$, with $i \in I^a$ \\
    
      \hskip 1.0cm \textbf{Predict} $\hat{Y}_{i}^{0} = \hat{g_{0}}(X_{i})$, with $i \in I^{a}$ \\
      \hskip 1.0cm \textbf{Predict} $\hat{Y}_{i}^{1} = \hat{g_{1}}(X_{i})$, with $i \in I^{a}$ \\ 
      \hskip 1.0cm \textbf{Predict} $\hat{D_{i}} = \hat{e}(X_{i})$, with $i \in I^{a}$ \\
	  \hskip 1.0cm \textbf{Predict} $\hat{Y_{i}} = \hat{\mu}(X_{i})$, with $i \in M$ \\   
        
        \hskip 1.5cm \textbf{Calculate} doubly-robust estimator (see equation \ref{equ:dr_calculation}) \\
        \hskip 1.5cm \textbf{Train} $\hat{S}_{i}$ = $l_{0}(X_i) + W $ with $i \in I^{a}$ \\
        \hskip 1.5cm \textbf{Predict} $\tilde{S}_{i} = \hat{l}(X_i)$ with $i \in M$ \\
        \hskip 2.0cm \textbf{Calculate} $\hat{V}_{i} = D_{i} - \hat{e}(X_i)$, with $i \in M$ \\
        \hskip 2.0cm \textbf{Calculate} $\hat{U}_{i} = Y_{i} - \hat{\mu}(X_i)$, with $i \in M$ \\
        \hskip 2.0cm \textbf{Store} $S_{b}^{*}(X)|i$ = $\tilde{S}_{i}(X)|b$  \\
        %\hskip 2.0cm \textbf{Store} $S_{b}(Z|i)$ for b = 1 to B
    \textbf{Estimate} GATES parameters ($\gamma$) with  OLS using $M$ (see equation \ref{equ:GATES_dr}) \\

      }
      \textbf{Average} $\gamma$ over $B$ iterations: $\Tilde{\gamma} = median\{\gamma\}$\\
      \textbf{Calculate} Density for every $i$: $\hat{\hat{S}}(X)|i$ given ${S}_{b}^{*}(X)|i$ over $b$  \\
      \textbf{Calculate} Final score-function ($\bar{S}_i(X)$) given density of medians for i = 1 to N  \\ 
    
    \caption{Extended GATES}
\end{algorithm}
\vspace{5mm}

The number of groups can, of course, be increased to e.g. 10. In empirical settings, it would depend on the sample size. If we want to have at least 30 observations within a group, we could have $\frac{N}{30 \times \Lambda}$ groups, with $\Lambda$-splits or folds of the dataset in the first stage used for training and testing. Here we consider only two-folds. However, there is no general relationship between the number of folds in cross-fitting and the precision of the estimator (see \cite{chernozhukov2018double} for an example with different folds).

\subsection{Alternative Grouping: The Baseline Effect}

In this paper, we focus on group average treatment effects that are based on quantiles from the CATE function. In the linear model we, therefore, condition on these quantiles to estimate the GATE parameters. The condition on which the sample is split into groups can, of course, be changed. Of particular interest could be the GATE based on different levels of the baseline outcome ($Y^0$). This has some similarity with the traditional ``Quantile Treatment Effect'', which is defined as $\delta(\lambda)=F_{Y^{1}}^{-1}(\lambda)-F_{Y^{0}}^{-1}(\lambda) \quad 0<\lambda<1$, where $F_{Y^{D}}^{-1}$ is the unconditional quantile function and $\lambda$ a specific quantile. Here, however, we only focus on $Y^0$. 

Let us think of $Y$ as being a binary variable, indicating if a customer bought something online or if a person participated in a program (e.g. unemployment training). We might believe that the treatment effect differs among different baseline probabilities. In such cases, we want to find group treatment effects based on quantiles of $Y^0$. As an estimator for $Y^0$, we can use the doubly-robust estimator and only concentrate on the baseline potential outcome: 
\begin{align}
\operatorname{E}&\left[{g}_{0}\left(X_{i} \right)+\frac{(1-D_{i})\left(Y_{i}-{g}_{0}\left(X_{i} \right)\right)}{ 1-{e_{0}}\left(X_{i}\right)} | X = x_i \right] = \operatorname{E}\left[Y^0 | X = x_i \right] 
\end{align}

The linear model would then have the following form:

\begin{align}
(Y-\mu_{0}(X))=  \sum_{k=1}^{K}\gamma_{k} \cdot (D-e_{0}(X)) \cdot \mathbf{I}({Y^0} \in I_k) + \nu. \label{equ:GATES_Y0}
\end{align}

\section{Simulation Study}

To evaluate the proposed extensions i) doubly-robust first stage, ii) orthogonal semi-parametric second stage, and iii) bagging of CATE, we use simulated data where the true treatment effects are known. In the following, we describe the data generating process (DGP) and show the variations that we consider. In all simulations we consider five groups denoted by $\gamma$. For example, $\gamma_1$ is the mean from the lower $\frac{100}{K}$\% quantile from the CATE function while $\gamma_K$ is the mean of the upper $\frac{100}{K}$\% quantile. Our method, which we call DO GATES (we refer to the name DO GATES as for Double Orthogonal GATES) estimates  $\gamma_k$ quantiles from a linear model. 
As a benchmark model, we use the mean-effect from each of the $K$-quantiles from the final treatment effect function ($\bar{S}(X)$) estimated via the doubly-robust estimator. Below, we show an example of the distribution from the estimated CATE function. A natural question is how the mean for the $K$-quantile groups, directly build from the CATE function, behaves without the second linear orthogonalization. To answer this, we compare the estimates from the benchmark and our approach with the mean values of $K = 5$ quantiles from the true treatment effect function. For the non-parametric estimation of functions, we use a random forest method (\textit{R-package: ranger}) as the corresponding machine learning algorithm due to its fast performance. 

In this simulation, we are interested in finite sample results and estimate our group treatment effects for $N = 2000$ and again for $N = 500$ observations.

\subsection{Data Generating Process}

The basic model used in this simulation study is a partially linear regression model based on \citet{robinson1988root}:

\begin{align}
Y = \tau_{0}(X)&D + \mu_{0}(X) + U,  &&E[U | X,D] = 0, \\
&D = e_{0}(X) + V,  &&E[V | X] = 0, \label{prop_score} \\
\tau(X) &= t_{0}(X) + W &&E[W| X], = 0, 
\end{align}

with $Y$ being a continuous  outcome variable. 
$\tau_{0}(X)$ is the true treatment effect or population uplift, while $D$ is the treatment status. The vector $X = ({X_{1},...,X_{p}})$ consists of $p$ different features, covariates or confounder.  $U$, $V$  and $W$ are unobserved covariates, which follow a random normal distribution = $N(0,1)$.  

Equation \ref{prop_score} is the propensity score. In the case of completely random treatment assignment, the propensity score $e_{0}(X_{i}) = c$ for all units ($i=1,...,N$). The scalar $c$ can take any value within the interval (0,1). In the simulation we consider $ c = 0.5$ (balanced) and $c = 0.8$ (imbalanced). 

%The covariates $X$ are generated from a random multivariate normal distribution ($N(0,1)$) %\footnote{Note that all values are continuous. In business applications, discrete values (categorical variables) are very common. For the data generation process as well as for the evaluation it would make no difference if such variables are present or not. This is since the used machine learning methods can handle categorical variables quite well. Exceptions are if there are many levels (e.g. the random forest package in R can only handle categorical variables with less than 56 different levels).} 
%as follows: \\

The function $\mu_{0}(X)$ takes the following form: 

\begin{align}
\mu_{0}(X) &=  X_{p/2} + X_{p/10} +  X_{p/4} \times X_{p/10}.
\end{align}
The vector $b = \frac{1}{l}$ with $l \in \{1,2,...,p\}$ represents weights for every covariate.

In the simulation, we focus on different functions of the \textbf{treatment assignment}. We use a CDF to create probabilities, which are then used in a Binomial function to create a binary treatment variable. The dependence of covariates within the normal distribution function is represented as $a(X)$, for which we use a variety of functions, namely random assignment with balanced and imbalanced groups, a linear dependence, interaction terms, and non-linear dependence. 
\begin{align}
&e_{0}(X) = \Phi\left(\frac{a(X)-\mu(a(X))}{\sigma(a(X))} \right),  \\
D \overset{ind.}{\sim} \text{Bernoulli}(&e_{0}(X)) \quad \text{such that} \quad D \in \{0;1\}.
\end{align}

\begin{align*}
\text{random assigment:} \quad &e_0(X) = c \quad \text{with} \quad c \in (0,1),\\
\text{linear:} \quad &a(X) = X_2 + X_{p/2} + X_{p/4} - X_{8},\\
\text{interaction:} \quad &a(X)  = X \times b + X_{p/2} + X_{2} + X_{p/4} \times X_{8}, \\
\text{non-linear:} \quad &a(X) = X \times b + \sin(X_{p/2}) + X_2 + \cos(X_{p/4} \times X_{8}).
\end{align*}

The \textbf{treatment effect} is heterogeneous and generated with a linear or a non-linear dependency of $X$:

\begin{align}
\text{linear:} \quad &\tau_{u}(X) = X_1 + X_2>0 + W \quad \text{with} \quad W \sim  N(0,0.5), \label{equ:tau_linear}\\
\text{non-linear:} \quad &\tau_{u}(X) = \sin(X \times b) + X_{5+p/2}. \label{equ:tau_non-linear}
\end{align}

We standardise the treatment effect within the interval $[0.1,1]$:
\begin{align}
\tau_{0}(X) &= \frac{\tau_{u}(X)-min(\tau_{u}(X))}{max(\tau_{u}(X))-min(\tau_{u}(X))}(1-0.1)+0.1.
\end{align}

Our aim is to perform a Monte Carlo simulation, in which we fix the distribution and dependence of the covariates (correlation). This is necessary to also fix the treatment effect function for each setting. All other functions can have small deviations within each repetition due to the error term and the random generation for the binomial distribution (for the treatment assignment).  The idea is to gather samples to approximate the distribution of the true treatment effect. Figure \ref{fig:density_theta} in the Appendix shows the true treatment effect function for the linear and non-linear case. They are quite normally distributed without heavy tails.

\subsection{Simulation Results}

In Table \ref{DGP_2000} and \ref{DGP_500}, we show the different scenarios, specifically for the treatment assignment mechanism. We also consider a simulation where both, the $\hat{e}(X)$ function as well as $\hat{g}_{D}(X_{i})$ (and as a consequence also $\hat{\mu}(X)$) is misspecified. We model misspecification by introducing an unobserved confounder in the DGP (namely $X_2$), which we exclude in the observed dataset. 
In scenarios $A$ to $F$, the data generation process for the treatment effect depends linear on the covariates, whereas in scenarios $G$ to $L$ the dependency is non-linear (see equations \ref{equ:tau_linear} and \ref{equ:tau_non-linear}). In Table \ref{DGP_2000} we consider N = 2000 observations. In Table \ref{DGP_500}, we set N =500. All other processes are the same. We repeat every DGP 100 times and estimate the mean absolute error (MAE) over all groups and repetitions. Another performance measurement is the squared bias as a comparison between the expected function and the true function. We aggregate our measurement as follows:

\begin{alignat}{3}
{}&MAE_{kj} &&= |\hat{\gamma}_k - \gamma_k|, \\
{}&MAE_{k} &&= \frac{1}{S}\sum_{j=1}^{S}\left(MAE_{kj}\right), \\
{}&MAE &&= \frac{1}{K}\sum_{k=1}^{K}\left(MAE_{k}\right).
\end{alignat}

\begin{alignat}{2}
&Bias^2(\hat{\gamma}_k) &&= (\underbrace{\mathbb{E}[\hat{\gamma}_k]}_{\frac{1}{S}\sum_{j=1}^{S}\hat{\gamma}_{kj}} - \gamma_k )^2 , \\
&Bias^2 &&= \frac{1}{K}\sum_{k=1}^{K}\left(Bias^2(\hat{\gamma}_k)\right).
\end{alignat}

\begin{table}[ht]
\resizebox{\textwidth}{!}{%
    \begin{threeparttable}
\caption{Settings and Monte Carlo averages for N = 2000.}
\label{DGP_2000}
\centering

\begin{tabular}{l|llllll}
\hline \hline \\
Scenarios & A/G      & B/H    & C/I        & D/J    & E/K   & F/L       \\ \hline
N         & 2000      & 2000   & 2000        & 2000     & 2000   & 2000       \\
$\mathbb{R}\bigsup{p}$       & 20       & 20     & 20         & 20       & 20     & 20         \\
$P(D=1)$ &$0.5 $ &0.2 (imbalanced)   &linear & interaction & non-linear & linear \\
Misspecification &No &No &No &No &No &Yes \\
MAE CATE 			&\textbf{0.06/0.05} 	&0.15/0.14 	&0.62/0.61 	&0.66/0.67 	&0.61/0.61 	&0.91/0.92  \\
MAE DO GATES 	&0.09/0.10 	&\textbf{0.08/0.10}	&\textbf{0.33/0.32}	&\textbf{0.32/0.36}	&\textbf{0.32/0.31} 	&\textbf{0.75/0.76} \\
$Bias^2$ CATE 		&\textbf{0.00}/0.00	&0.03/0.03	&0.47/0.44	&0.49/0.50	&0.46/0.44	&0.99/1.00 \\
$Bias^2$ DO GATES &0.01/0.00				&\textbf{0.00}/\textbf{0.00}	&\textbf{0.15}/\textbf{0.15}	&\textbf{0.17}/\textbf{0.18}	&\textbf{0.15}/\textbf{0.13}	&\textbf{0.74}/\textbf{0.75} \\
\hline \hline
\end{tabular}
\begin{tablenotes}
      \small
      \item \textit{Notes:} Mean absolute error (MAE) of the average treatment effect (estimated from the quantile estimates)  over S = 100 Monte Carlo simulations. Setting G to L consists of a non-linear treatment effect function.
    \end{tablenotes}
  \end{threeparttable}
}
\end{table}

\begin{figure}[ht!]
\begin{center}
\includegraphics[width=0.86\textwidth]{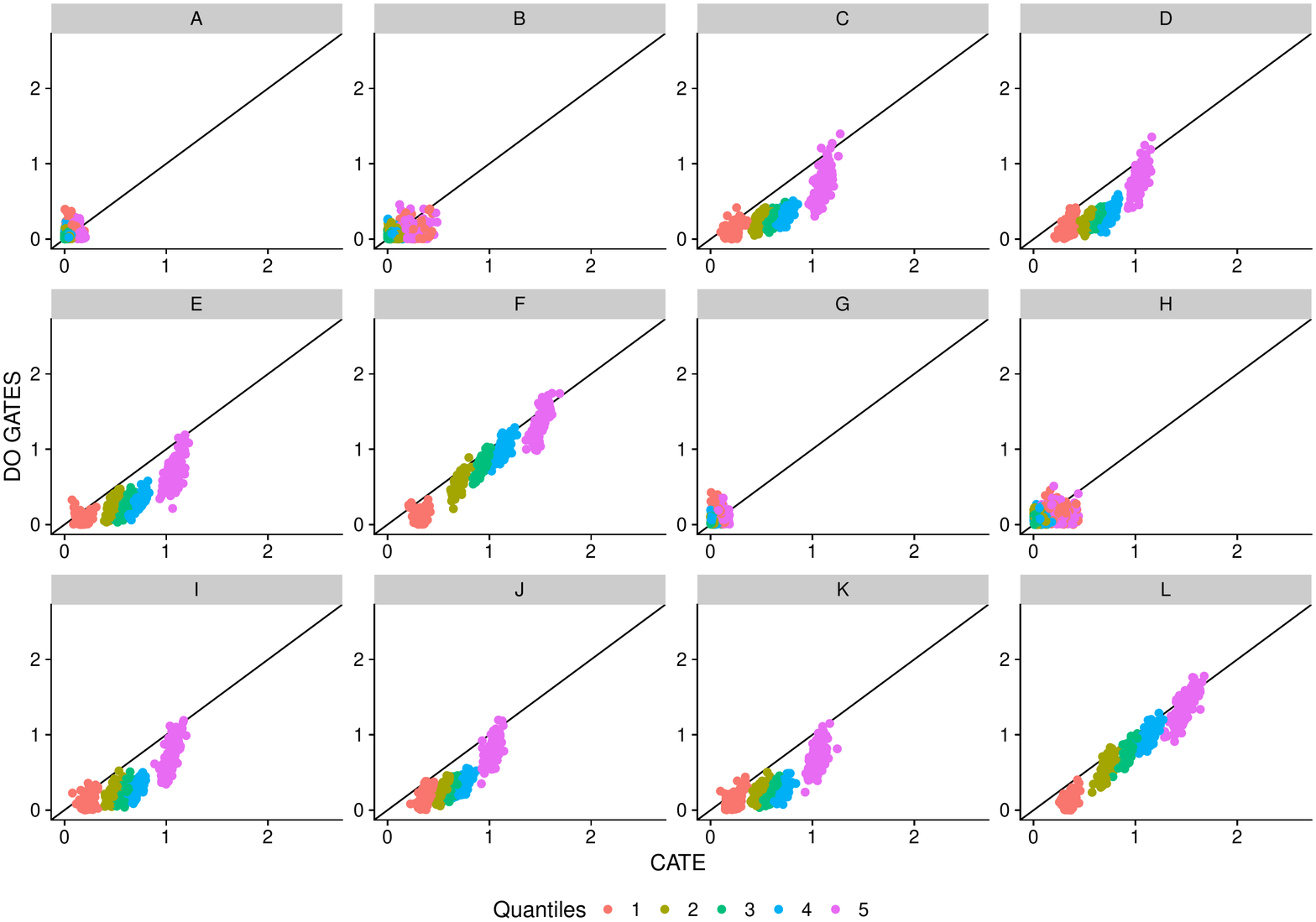}
\caption{Group estimate from the CATE function and DO GATES. Groups are coloured by quantiles (1 = least affected, 5 = most affected). Axes show the absolute error between estimates and true quantile treatment effects. The 45-degree line indicates the equality of both methods. Th number of observations equals 2000.}
\label{fig:cate_vs_dr_2000}
\end{center}
\end{figure}

\begin{figure}[ht!]
\begin{center}
\includegraphics[width=0.86\textwidth]{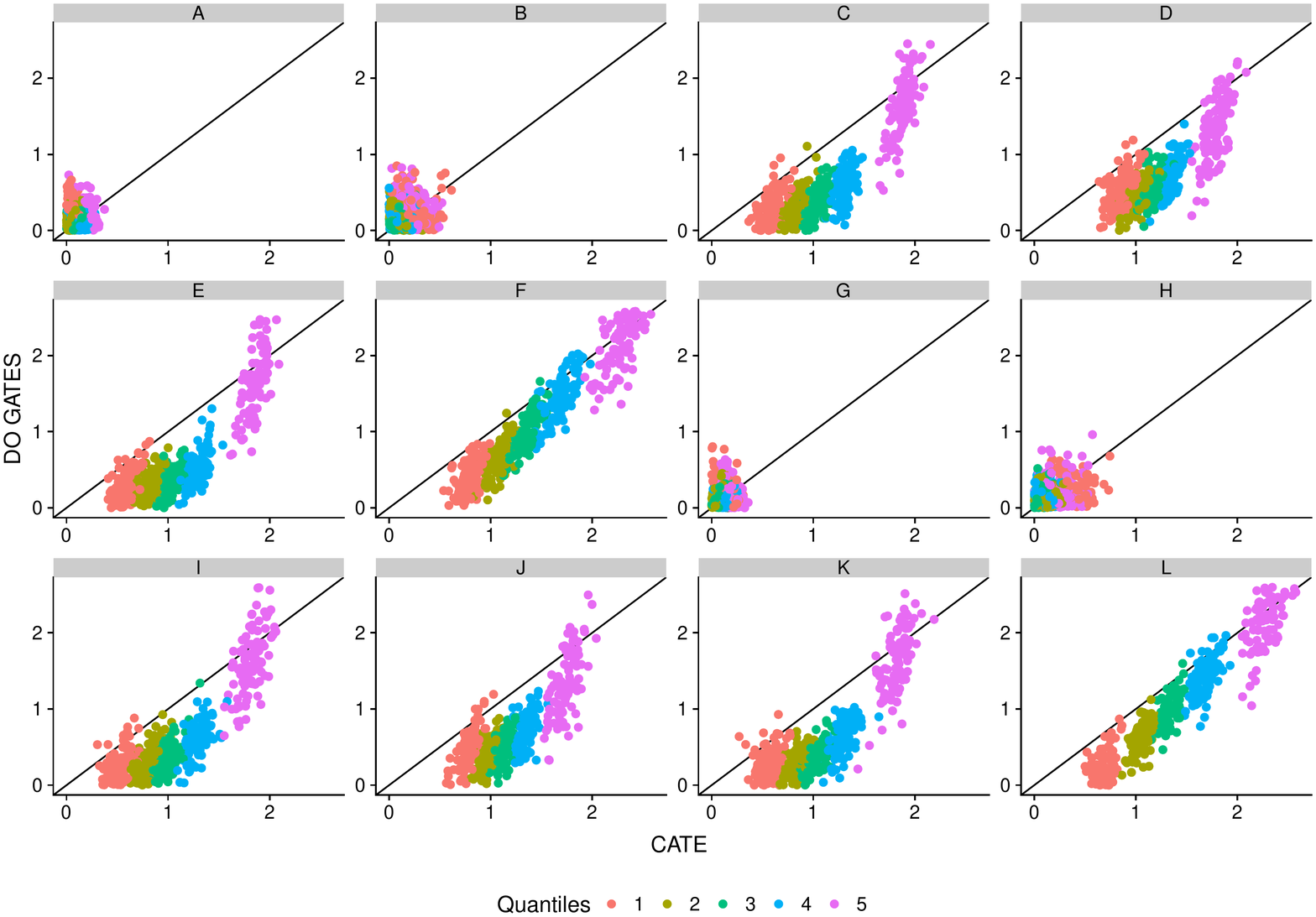}
\caption{Group estimate from the CATE function and DO GATES. Groups are coloured by quantiles (1 = least affected, 5 = most affected). Axes show the absolute error between estimates and true quantile treatment effects. The 45-degree line indicates the equality of both methods. Th number of observations equals 500.}
\label{fig:cate_vs_dr_500}
\end{center}
\end{figure}

In Figure \ref{fig:cate_vs_dr_2000} and \ref{fig:cate_vs_dr_500} we plot the results for each group of the simulation study for N = 2000 and 500 observations, respectively. We find that especially the lower quantiles show a smaller absolute error of the quantile treatment effect. If the treatment assignment mechanism is randomized, our model performs competitively or slightly worse than the doubly-robust estimator, as can be seen from scenario A, B, G and H. In all observational settings, we find a smaller mean absolute error from our DO GATES method compared to the benchmark model. We find that the highest error from the true treatment effect arises when both models are misspecified as in scenario F and L. There is, however, no significant difference in the MAE if we change the dependency of the treatment effect from linear to non-linear. 

The bias behaves like the MAE in the sense that only for the balanced randomization the two models show equal bias. In all other settings, we find a smaller bias for our method and again the highest differences when there is selection-bias, which depends linear and or with interactions on the treatment. Table \ref{AE_Groups_N2000} and \ref{AE_Groups_N500} in the Appendix show the MAE for each group and DGP over all repetitions. We also notice that the deviation between the two methods is highest for the lowest group and decreases towards the upper quantiles. When we set N = 500, we find that the MAE and the BIAS increase in both methods but the overall structure stays the same as for N = 2000.

%\clearpage
%\newpage

\begin{table}[ht]
\resizebox{\textwidth}{!}{%
    \begin{threeparttable}
\caption{Settings and Monte Carlo averages for N = 500.}
\label{DGP_500}
\centering
\begin{tabular}{l|llllll}
\hline \hline \\
Scenarios & A/G      & B/H    & C/I        & D/J    & E/K   & F/L       \\ \hline
N         & 500      & 500   & 500        & 500     & 500   & 500       \\
$\mathbb{R}\bigsup{p}$       & 20       & 20     & 20         & 20       & 20     & 20         \\
$P(D=1)$ &$0.5 $ &0.2 (imbalanced)   &linear & interaction & non-linear & linear \\
Misspecification &No &No &No &No &No &Yes \\
MAE CATE 			&\textbf{0.12/}\textbf{0.14 }	&\textbf{0.16}/0.20 	&1.13/1.10		&1.22/1.21 	&1.11/1.10 	&1.44/1.41   \\
MAE DO GATES 	&0.19/0.16 	&0.21/\textbf{0.19}	&\textbf{0.64}/\textbf{0.63}		&\textbf{0.71}/\textbf{0.69} 	&\textbf{0.62}/\textbf{0.64} 	&\textbf{1.13}/\textbf{1.10} \\
$Bias^2$ CATE 		&\textbf{0.02}/0.02	&0.03/0.05	&1.47/1.38	&1.60/1.54	&1.43/1.40	&2.34/2.29 \\
$Bias^2$ DO GATES &0.03/0.02				&\textbf{0.02}/\textbf{0.00}	&\textbf{0.67}/\textbf{0.65}	&\textbf{0.60}/\textbf{0.59}	&\textbf{0.63}/\textbf{0.66}	&\textbf{1.67}/\textbf{1.69} \\
\hline \hline
\end{tabular}
\begin{tablenotes}
      \small
      \item \textit{Notes:} Mean absolute error (MAE) of the average treatment effect (estimated from the quantile estimates)  over S = 100 Monte Carlo simulations.  Setting G to L consist of a non-linear treatment effect function.
    \end{tablenotes}
  \end{threeparttable}
}
\end{table}

\subsection{Model Averaging of CATE}

\begin{figure}[ht]
\begin{center}
\includegraphics[width=0.9\textwidth]{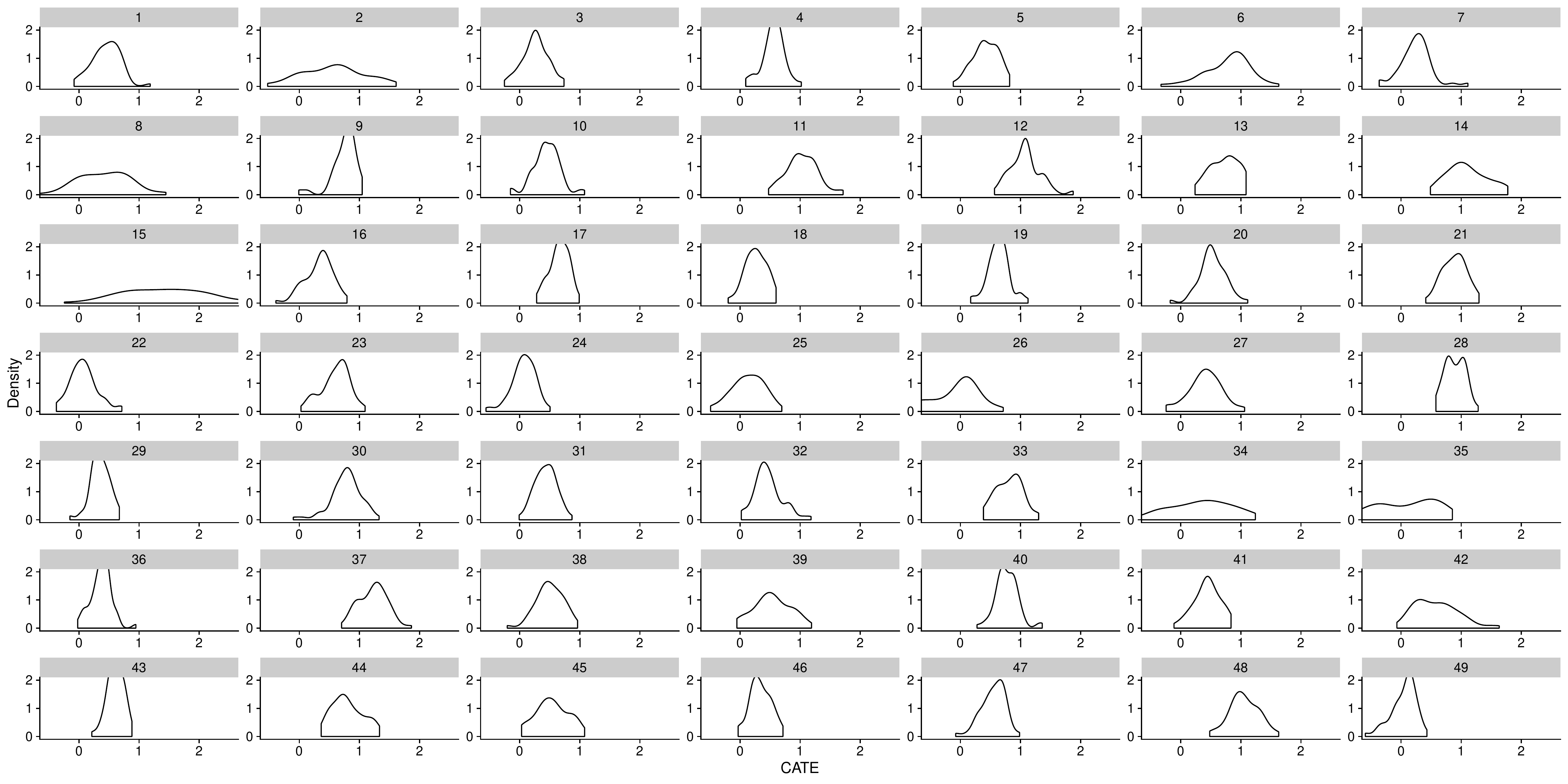}
\caption{Distribution of score function ($\hat{\hat{S}}_b(X)$) for each of 49 randomly selected individuals. Density is estimated over 100 iterations. Due to random sample splitting, the number of estimates for each individual is between 39 and 59 for 100 iterations. }
\label{fig:density}
\end{center}
\end{figure}

Figure \ref{fig:density} shows the estimates from the conditional average treatment effect over $B = 100$ bootstrap iterations (we split the sample once in equal parts). For every iteration, we either get an estimate for a specific observation $i$ if this observation is in the test-sample or we don't which we label with a ``NA'' estimate. After 100 iterations we have at least 39 estimates for each observation. If the specific sample we use for training would always be equal in terms of distributions, we would assume that every iteration estimates the same value for a specific observation. Here we show that this is not the case since the sample splitting does matter, especially in finite sample settings.

The simulated data, in this case, has the following properties. $N = 1000$, $X = \mathbb{R}^{20}$, $e_{0}(X) = 0.5$ and $\tau_{u}(X) = X_1 + X_2>0 + W \quad \text{with} \quad W \sim  N(0,0.5)$. We standardize $\tau_u$ to $\tau(X) \in [0.1,1]$. The densities for 49 selected observations show that even in randomised experiments, the point estimates differ due to the sample splitting in the first step. We propose to average the estimates by taking the median over each iteration. This leads to a more stable conditional treatment effect function: 

\begin{align}
\bar{S}_i(X) = median\{\hat{\hat{S}}_b(X)\}
\end{align}

\begin{figure}[ht]
\begin{center}
\includegraphics[width=0.9\textwidth]{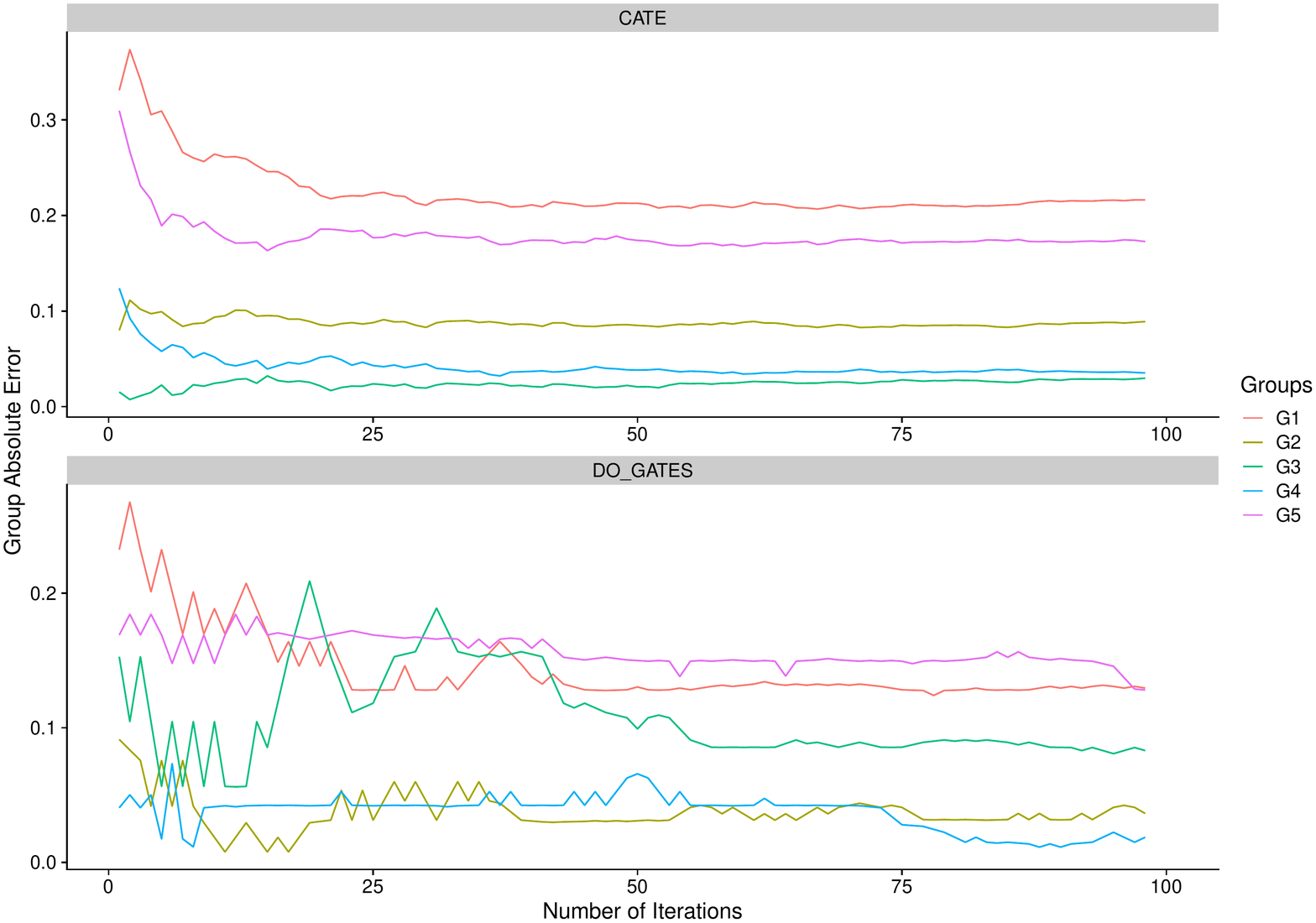}
\caption{Upper plot shows the absolute error for the CATE estimator given $B$ iterations. Lower plot shows the same for the DO GATES method. }
\label{fig:AE_B_iterations}
\end{center}
\end{figure}

We show in Figure \ref{fig:AE_B_iterations} how the absolute error (AE) for each group depends on the number of iterations. The upper plot shows the AE for the CATE function and the lower plot for the DO GATES method. The AE for the CATE function stabilizes after around 50 iterations while for the DO GATES we can decrease the AE further as we take the median over more iterations. Interestingly, it seems that the variance in AE for the DO GATES method does not mainly depend on the CATE function as we can see after taking the median of 50 iterations. The CATE function is quite stable but the GATES still show some variance in the AE. It could be that this is caused in the orthogonalization step where we train the two conditional mean functions. We also try to average with the arithmetic mean and find no significant difference between the two approaches. However, it could be that in some scenarios, like a very small sample size, single estimates are far away from the true estimate and should be treated as outliers. Therefore, we suggest to use the median for averaging.

\clearpage

\section{Conclusion}

In this paper, we propose a method to estimate group average treatment effects using machine learning methods and linear estimation for non-randomized control trials. Since flexibility in terms of the model choice, as well as interpretability of the results, is our main interest, we extend the idea of the GATES approach by using a doubly-robust estimator in the first, the non-parametric, step. In the second step, the linear projection function, we use Neyman-orthogonal moments to overcome the regularization bias due to the dependency of the covariates on the outcome and the treatment assignment parameter. This ensures to control for self-selection into treatment which is a realistic challenge in observational studies. Our main interest is to find heterogeneity in terms of quantiles from the treatment effect function and estimate them via a best linear projection. We also propose to weight the CATE function by taking the median over $B$ iterations instead of single estimates. Since we are interested in describing the groups in terms of all observations, random sample splitting in half of the sample only gives us estimates from half of the observations. After approximately $B=10$ iterations we have at least one estimate for each observation. Therefore, we suggest to use at least 50 or preferably 100 iterations in order to cover the whole sample. 

We find that our proposed model has a lower MAE and a lower Bias for the majority of the groups we consider and among different simulation scenarios, compared to the benchmark model. In randomized control trials, both methods perform competitive. As soon as we include selection-bias, the DO GATES method performs better due to its second orthogonalization step. We note that a natural extension is to estimate the ATE by taking the mean over the group estimates. However, this estimator is not the scope of the paper and there are direct approaches to estimate the ATE. 
%This amount needs to be increased to e.g. 50, respectively. 
%So far we only consider a random forest algorithm (using the ranger R-package for fast computation) as a machine learning method. 
%We will extend this to the use of boosted gradient descent (XGBoost), neural networks and linear methods like variants of the Elastic Net. We can also consider different models for each nuisance function. 

%%%%%%%%%%%%%%%%%%%%%%%%%%%%%%%%%%%%%%%%%%%%%%%%%%%%%%%%%%%
%%%%%%%%%%%%%%%%%%%%%%%%%%%%%%%%%%%%%%%%%%%%%%%%%%%%%%%%%%%

% ----------------
% --- appendix ---
% ----------------
\appendix

% literature
\FloatBarrier % Dadurch wird dieser Abschnitt nicht unterbrochen
\newpage

\phantomsection \addcontentsline{toc}{section}{References}
% define citation style
\bibliographystyle{plainnat}
\bibliography{literature}

%%%%%%%%%%%%%%%%%%%%%%%%%%%%%%%%%%%%%%%%%%%%%%%%%%%%%%%%%%%
%%%%%%%%%%%%%%%%%%%%%%%%%%%%%%%%%%%%%%%%%%%%%%%%%%%%%%%%%%%
% figures

\clearpage
\section{Figures}

\begin{figure}[ht!]
\begin{center}
\includegraphics[width=0.9\textwidth]{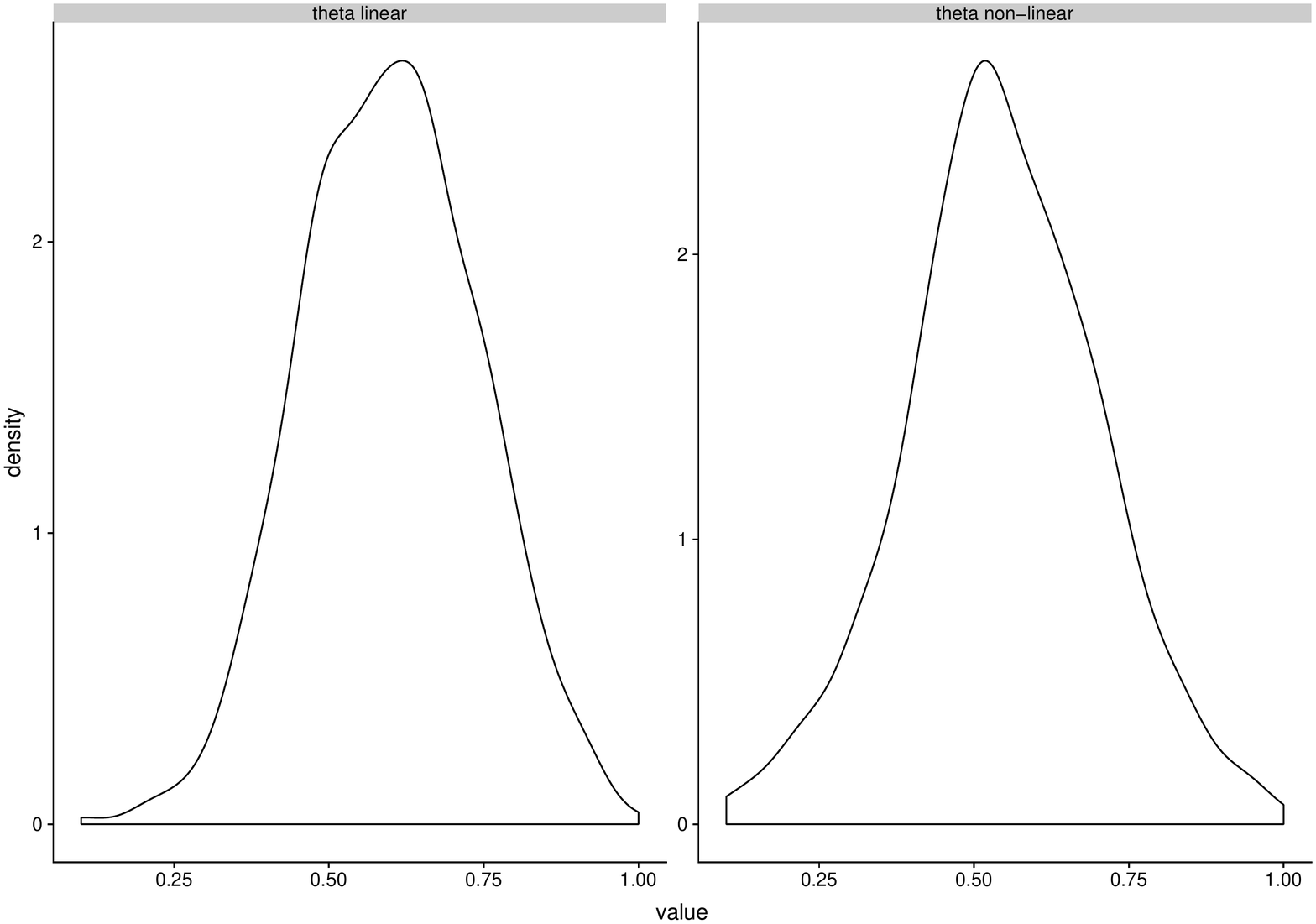}
\caption{Density for the true and fixed heterogeneous treatment effect function. Left plot shows the linear treatment effect dependency and right plot the non-linear dependency. }
\label{fig:density_theta}
\end{center}
\end{figure}

%%%%%%%%%%%%%%%%%%%%%%%%%%%%%%%%%%%%%%%%%%%%%%%%%%%%%%%%%%%
%%%%%%%%%%%%%%%%%%%%%%%%%%%%%%%%%%%%%%%%%%%%%%%%%%%%%%%%%%%
% tables (not mandatory)
\clearpage

\section{Tables}

\begin{table}[ht]
\resizebox{0.9\textwidth}{!}{%
    \begin{threeparttable}
\caption{Mean absolute error by setting and groups for N = 2000.}
\label{AE_Groups_N2000}
\centering
\begin{tabular}{ccrr|ccrr}
  \hline
  SETTING & G & DO GATES & CATE & SETTING & G & DO GATES & CATE \\ 
  \hline \hline
 A & 1 & 0.18 & \textbf{0.07} & G & 1 & 0.18 & \textbf{0.07} \\ 
 A & 2 & 0.07 & \textbf{0.03} & G & 2 & 0.08 & \textbf{0.02} \\ 
 A & 3 & 0.04 & \textbf{0.02} & G & 3 & 0.05 & \textbf{0.02} \\ 
 A & 4 & 0.07 & \textbf{0.03} & G & 4 & 0.07 & \textbf{0.04} \\ 
 A & 5 & \textbf{0.12} & 0.13 & G & 5 & 0.14 & \textbf{0.12} \\ 
 B & 1 & \textbf{0.12} & 0.27 & H & 1 & \textbf{0.15} & 0.25 \\ 
 B & 2 & 0.07 & 0.07 & H & 2 & 0.08 & \textbf{0.07} \\ 
 B & 3 & 0.06 & \textbf{0.03} & H & 3 & 0.06 & \textbf{0.04} \\ 
 B & 4 & \textbf{0.07} & 0.11 & H & 4 & \textbf{0.08} & 0.09 \\ 
 B & 5 & \textbf{0.12} & 0.29 & H & 5 & \textbf{0.14} & 0.27 \\ 
 C & 1 &\textbf{ 0.11} & 0.21 & I & 1 & \textbf{0.12} & 0.23 \\ 
 C & 2 & \textbf{0.25} & 0.49 & I & 2 & \textbf{0.23} & 0.48 \\ 
 C & 3 & \textbf{0.25} & 0.61 & I & 3 & \textbf{0.23} & 0.59 \\ 
 C & 4 & \textbf{0.30} & 0.74 & I & 4 & \textbf{0.29} & 0.71 \\ 
 C & 5 & \textbf{0.73} & 1.10 & I & 5 & \textbf{0.73} & 1.05 \\ 
 D & 1 & \textbf{0.16} & 0.34 & J & 1 & \textbf{0.16} & 0.37 \\ 
 D & 2 & \textbf{0.23} & 0.53 & J & 2 & \textbf{0.25} & 0.55 \\ 
 D & 3 & \textbf{0.27} & 0.64 & J & 3 & \textbf{0.30} & 0.66 \\ 
 D & 4 & \textbf{0.35} & 0.75 & J & 4 & \textbf{0.35} & 0.75 \\ 
 D & 5 & \textbf{0.78} & 1.03 & J & 5 & \textbf{0.77} & 1.03 \\ 
 E & 1 & \textbf{0.09} & 0.18 & K & 1 & \textbf{0.12} & 0.23 \\ 
 E & 2 & \textbf{0.25} & 0.48 & K & 2 & \textbf{0.24} & 0.48 \\ 
 E & 3 & \textbf{0.26} & 0.60 & K & 3 & \textbf{0.23} & 0.59 \\ 
 E & 4 & \textbf{0.30} & 0.72 & K & 4 & \textbf{0.27} & 0.70 \\ 
 E & 5 & \textbf{0.73} & 1.08 & K & 5 & \textbf{0.70} & 1.05 \\ 
 F & 1 & \textbf{0.10} & 0.32 & L & 1 & \textbf{0.11} & 0.36 \\ 
 F & 2 & \textbf{0.54} & 0.70 & L & 2 & \textbf{0.56} & 0.72 \\ 
 F & 3 & \textbf{0.78} & 0.92 & L & 3 & \textbf{0.78} & 0.93 \\ 
 F & 4 & \textbf{1.00} & 1.14 & L & 4 &\textbf{ 0.99} & 1.13 \\ 
 F & 5 & \textbf{1.36} & 1.50 & L & 5 & \textbf{1.36} & 1.49 \\ 

   \hline \hline
\end{tabular}
\begin{tablenotes}
      \small
      \item \textit{Notes:} Settings as in Table \ref{DGP_2000}. Number of observations: N = 2000. Averages over 100 repetitions. 
    \end{tablenotes}
  \end{threeparttable}
}
\end{table}

\begin{table}[ht]
\resizebox{0.9\textwidth}{!}{%
    \begin{threeparttable}
\caption{Mean absolute error by setting and groups for N = 500.}
\label{AE_Groups_N500}
\centering
\begin{tabular}{ccrr|ccrr}
  \hline \hline
 SETTING & G & DO GATES & CATE & SETTING & G & DO GATES & CATE \\ 
  \hline
 A & 1 & 0.34 & \textbf{0.08} & G & 1 & 0.27 & \textbf{0.14} \\ 
 A & 2 & 0.16 & \textbf{0.05} & G & 2 & 0.16 & \textbf{0.06} \\ 
 A & 3 & 0.12 & \textbf{0.09} & G & 3 & 0.12 & \textbf{0.10} \\ 
 A & 4 & \textbf{0.11 }& 0.14 & G & 4 & \textbf{0.09} & 0.15 \\ 
 A & 5 & \textbf{0.21} & 0.23 & G & 5 & \textbf{0.19} & 0.23 \\ 
 B & 1 & 0.29 & 0.29 & H & 1 & \textbf{0.25} & 0.39 \\ 
 B & 2 & \textbf{0.18} & 0.10 & H & 2 & 0.15 & \textbf{0.13} \\ 
 B & 3 & 0.13 & \textbf{0.07} & H & 3 & 0.14 & \textbf{0.07} \\ 
 B & 4 & 0.17 & \textbf{0.10} & H & 4 & 0.16 & \textbf{0.13} \\ 
 B & 5 & 0.30 & \textbf{0.25} & H & 5 & \textbf{0.28} & 0.29 \\ 
 C & 1 & \textbf{0.31} & 0.60 & I & 1 & \textbf{0.27} & 0.52 \\ 
 C & 2 & \textbf{0.31} & 0.83 & I & 2 & \textbf{0.31} & 0.80 \\ 
 C & 3 & \textbf{0.37} & 1.04 & I & 3 & \textbf{0.37} & 1.01 \\ 
 C & 4 & \textbf{0.60} & 1.31 & I & 4 & \textbf{0.59} & 1.29 \\ 
 C & 5 & \textbf{1.64} & 1.88 & I & 5 & \textbf{1.63} & 1.82 \\ 
 D & 1 & \textbf{0.55} & 0.85 & J & 1 & \textbf{0.47} & 0.79 \\ 
 D & 2 & \textbf{0.44} & 1.00 & J & 2 & \textbf{0.43} & 0.98 \\ 
 D & 3 & \textbf{0.53} & 1.14 & J & 3 & \textbf{0.50} & 1.13 \\ 
 D & 4 & \textbf{0.73} & 1.35 & J & 4 & \textbf{0.73} & 1.34 \\ 
 D & 5 & \textbf{1.30} & 1.78 & J & 5 & \textbf{1.35} & 1.74 \\ 
 E & 1 & \textbf{0.31} & 0.58 & K & 1 & \textbf{0.29} & 0.55 \\ 
 E & 2 & \textbf{0.31} & 0.81 & K & 2 & \textbf{0.29} & 0.82 \\ 
 E & 3 & \textbf{0.33} & 1.02 & K & 3 & \textbf{0.37} & 1.03 \\ 
 E & 4 & \textbf{0.53} & 1.29 & K & 4 & \textbf{0.62} & 1.30 \\ 
 E & 5 & \textbf{1.62} & 1.86 & K & 5 & \textbf{1.64} & 1.82 \\ 
 F & 1 & \textbf{0.38} & 0.79 & L & 1 & \textbf{0.23} & 0.69 \\ 
 F & 2 & \textbf{0.65} & 1.09 & L & 2 & \textbf{0.62} & 1.06 \\ 
 F & 3 & \textbf{0.97} & 1.35 & L & 3 & \textbf{0.99} & 1.33 \\ 
 F & 4 & \textbf{1.42} & 1.68 & L & 4 & \textbf{1.43} & 1.68 \\ 
 F & 5 & \textbf{2.21} & 2.30 & L & 5 & \textbf{2.25} & 2.31 \\ 

   \hline \hline
\end{tabular}
\begin{tablenotes}
      \small
      \item \textit{Notes:} Settings as in Table \ref{DGP_500}. Number of observations: N = 500. Averages over 100 repetitions. 
    \end{tablenotes}
  \end{threeparttable}
}
\end{table}

\clearpage
\addtocontents{toc}{\protect\enlargethispage{1\normalbaselineskip}}

\section{Technical Appendix}

\subsection{Correlated Covariates}

\begin{figure}[ht]
\begin{center}
\includegraphics[scale=0.44]{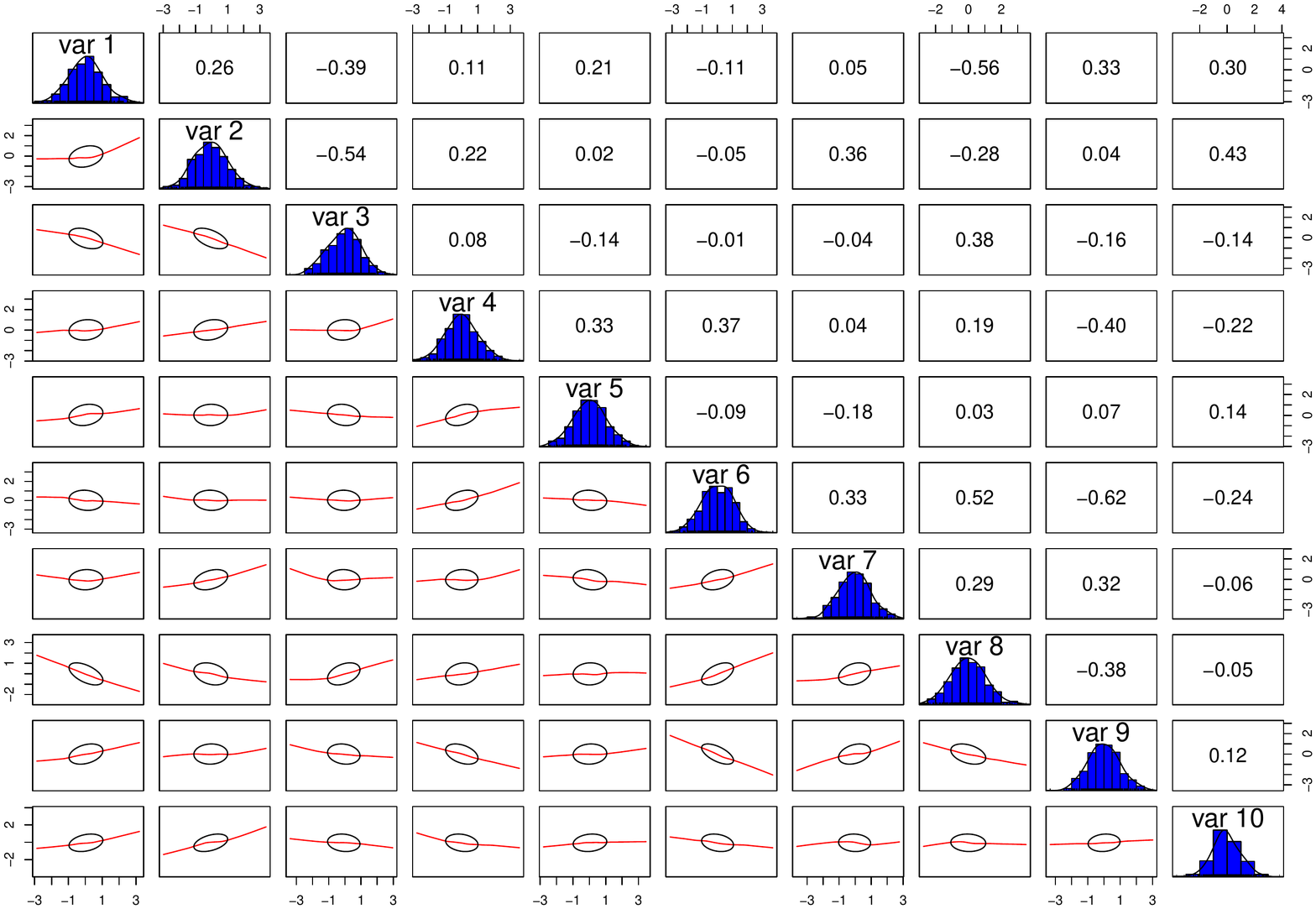}
\caption[Correlation Matrix of Covariates.]{Correlation Matrix of Covariates. Correlation metric is Bravais-Pearson.}
\label{Cov_matrix}
\end{center}
\end{figure}

We show how we generate the covariates and their correlation. We assume correlation through a uniform distribution of the covariance matrix, which is then transformed to a correlation matrix. Correlated chararcteristics are more common in real datasets and helps to investigate the performance of ML algorithms, especially the regularization bias, in a more realistic manner. Figure \ref{Cov_matrix} shows the correlation matrix for 10 randomly selected covariates to give an example of the correlation.

\vspace{5mm}

\begin{algorithm}[H] \label{Cov_Generation}
    %\SetKwInOut{Input}{Input}
%\small
    
  \textbf{Generate} random positive definite covariance matrix $\Sigma$ based on a uniform distribution over the space $k \times k$ of the correlation matrix. \\
  \textbf{Scale covariance matrix} This equals the correlation matrix and can be seen as the covariance matrix of the standardized random variables $\Sigma = \frac{X}{\sigma(X)}$. \\
  \textbf{Generate} random normal distributed variables $X_{N \times k}$ with mean = 0 and variance = $\Sigma$.\\
    
    \caption{Generation of Covariates}
\end{algorithm}

\vspace{5mm}

%\subsection{Supplementary Analysis}

\end{document}